\documentclass[square]{article}

    \PassOptionsToPackage{numbers, compress}{natbib}

\usepackage[final]{neurips_2024}
\usepackage{threeparttable}

\usepackage[utf8]{inputenc} 
\usepackage[T1]{fontenc}    
\usepackage{hyperref}       
\usepackage{url}            
\usepackage{booktabs}       
\usepackage{amsfonts}       
\usepackage{nicefrac}       
\usepackage{microtype}      
\usepackage{xcolor}         
\usepackage{graphicx}
\usepackage{multirow}
\usepackage{algorithm}
\usepackage{algorithmic}

\title{\textit{CoVoMix}: Advancing Zero-Shot Speech Generation for Human-like Multi-talker Conversations}

\author{%
  \small{Leying Zhang}$^{1,2}$\thanks{Work done during an internship at Microsoft Azure AI. zhangleying@sjtu.edu.cn} 
  \And \small{Yao Qian}$^{2}$ \thanks{Correspondence: yaoqian@microsoft.com} 
  \And \small{Long Zhou}$^{2}$ 
  \And \small{Shujie Liu}$^{2}$
  \And \small{Dongmei Wang}$^{2}$
  \And \small{Xiaofei Wang}$^{2}$
  \And \small{Midia Yousefi}$^{2}$
  \And \small{Yanmin Qian}$^{1}$
  \And \small{Jinyu Li}$^{2}$
  \And \small{Lei He}$^{2}$
  \And \small{Sheng Zhao}$^{2}$
  \And \small{Michael Zeng}$^{2}$
  \And
  $^{1}$\texttt{Shanghai Jiao Tong University,  China} 
  \And
  $^{2}$\texttt{Microsoft, USA} 
}

\begin{document}

\maketitle

\begin{abstract}
Recent advancements in zero-shot text-to-speech (TTS) modeling have led to significant strides in generating high-fidelity and diverse speech. However, dialogue generation, along with achieving human-like naturalness in speech, continues to be a challenge. In this paper, we introduce CoVoMix: \textbf{Co}nversational \textbf{Vo}ice \textbf{Mix}ture Generation, a novel model for zero-shot, human-like, multi-speaker, multi-round dialogue speech generation. CoVoMix first converts dialogue text into multiple streams of discrete tokens, with each token stream representing semantic information for individual talkers. These token streams are then fed into a flow-matching based acoustic model to generate mixed mel-spectrograms. Finally, the speech waveforms are produced using a HiFi-GAN model. Furthermore, we devise a comprehensive set of metrics for measuring the effectiveness of dialogue modeling and generation. Our experimental results show that CoVoMix can generate dialogues that are not only human-like in their naturalness and coherence but also involve multiple talkers engaging in multiple rounds of conversation. This is exemplified by instances generated in a single channel where one speaker's utterance is seamlessly mixed with another's interjections or laughter, indicating the latter's role as an attentive listener. Audio samples are available at https://aka.ms/covomix. 
\end{abstract}

\begin{figure}[h]
  \centering
\centerline{\includegraphics[width=1.0\linewidth]{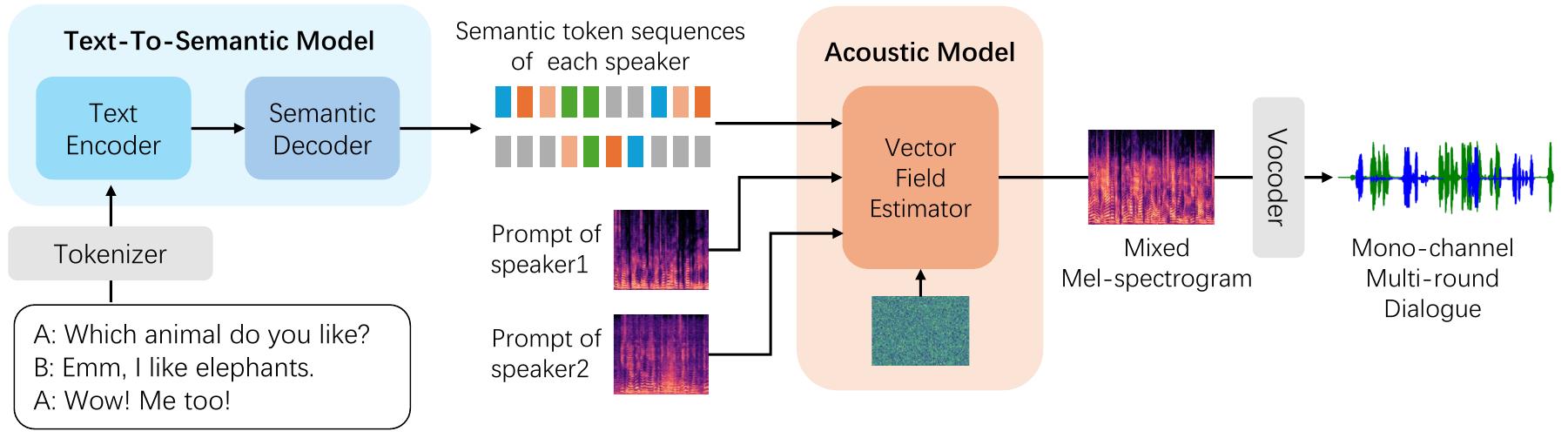}}
\caption{The overview of CoVoMix framework, which consists of a multi-stream text-to-semantic model, a conditional flow-matching based acoustic model for mixed mel-spectrogram generation, and a HiFi-GAN based vocoder for waveform production.
}
\label{fig:intro}
\end{figure}

\section{Introduction}
Zero-shot Text-to-speech (TTS) technology aims to create human-like natural speech with voice characteristics prompted by the context. Recent deep learning advancements have significantly improved synthesized speech quality, especially in formal reading scenarios~\cite {tan2021survey, wang2023neural, shen2023naturalspeech}. However, TTS systems still struggle with rendering spontaneous-style speech and managing seamless transitions during conversations—common occurrences in everyday human communication~\cite{li23ba_interspeech,levinson2020human}. On the one hand, spontaneous-style speech encompasses phenomena like filled pauses, interjections, repairs, repetitions, and laughter, which lend human-realistic naturalness to spoken language~\cite{ward1989understanding}. On the other hand, in natural conversations, speakers intuitively time their speech, determining when to speak and when to yield, resulting in seamless transitions with appropriate overlaps or moments of silence~\cite{SCHEGLOFF_2000,nguyen2023generative}. Overlapping speech, defined as more than one person is speaking, can easily exceed 20\%, in informal gathering conversational speech~\cite{8682572}. Considering these conversational features, we summarize three main challenges in generating spontaneous dialogues.

First, the scarcity of high-quality, spontaneous conversational datasets, along with the difficulty in segmenting paralinguistic behaviors, continues to be a significant obstacle in the field. Spontaneous behavior and non-verbal expressions such as laughter, receive insufficient attention in speech synthesis. Existing high-quality datasets for spontaneous and conversational speech are relatively small and involve a limited number of speakers~\cite{lee2023dailytalk}. Identifying and segmenting these paralinguistic features, as highlighted by studies~\cite{nagata2018defining, tits20b_interspeech}, poses difficulties. Models that require pre-alignment necessitate sophisticated manual annotation. Otherwise, the low-quality data can adversely impact performance, particularly in TTS tasks~\cite{yu2023autoprep}. 

Second, research on turn-taking mechanisms in multi-speaker dialogues is less explored. In such dialogues, the forthcoming speaker anticipates the end of the current speaker's turn by analyzing structural and contextual cues, and then begins their speech seamlessly at the anticipated transition point~\cite{sacks1978simplest}. Speakers tend to adapt the pause length to match other participants~\cite{heldner2010pauses}. Overlapping speech occurs when one speaker starts talking before another finishes, which can be a sign of enthusiasm or an attempt to take turns~\cite{dethlefs2016information,zhang2022enroll,zhang2024ddtse}.  

Third, the consistency in multi-round dialogues is not guaranteed in conventional methods. Simply concatenating each utterance to form a dialogue may result in inconsistent speaker characteristics, particularly when the same speaker engages in multi-round dialogue. In addition, the context of the preceding utterance plays an important role in the control of pauses and prosody, and thus influences the naturalness of generated dialogues~\cite{heldner2010pitch}.

To effectively generate human-like dialogue, we propose CoVoMix, named Conversational Voice Mixture Generation, for multi-talker dialogue generation, shown in Figure~\ref{fig:intro}. 
The main contributions of the paper can be summarized as follows: 
\begin{enumerate}
    \item To the best of our knowledge, it is the first attempt at zero-shot, human-like, multi-talker conversational mixed speech generation. We propose 1) a simultaneous multi-stream semantic token prediction, with each stream representing an individual talker, from dialogue text; and 2) a multi-talker flow-matching based acoustic model for generating a mixed mono mel-spectrogram given multiple contexts. It is capable of generating single-channel multi-round dialogue containing multiple speakers concurrently, enabling simultaneous timbre cloning of multiple speakers in zero-shot scenarios.


    \item  We design a variety of evaluation metrics for dialogue generation, and demonstrate that the CoVoMix model is proficient at generating both human-like dialogues and monologues, exhibiting natural speaker turn-taking, realistic vocal burst-like laughter, consistent speech in terms of speaker similarity throughout multiple rounds of dialogue.

   \item We employ the Fisher dataset~\cite{cieri2004fisher} for this study, which was curated for robust speech recognition. Our approach includes a comprehensive strategy for processing this dataset, including both training and evaluation for monologue and dialogue speech. The data processing script, along with the model training and inference codes are publicly available~\footnote{https://github.com/vivian556123/NeurIPS2024-CoVoMix.git}.

\end{enumerate}

\section{Related Work}

\subsection{Zero-shot Text-to-Speech }
The goal of Zero-shot TTS is to synthesize speech in a target voice which was unseen during training, given only the target transcript and a short reference of the target voice ~\cite{tan2021survey,taylor2009text,mu2021review, casanova2022yourtts,lengprompttts}
. Zero-shot TTS systems are generally divided into two categories: \emph{(i)} Diffusion-based Zero-Shot TTS~\cite{shen2023naturalspeech, le2024voicebox, huang2022fastdiff, huang2022prodiff, jeong2021diff, kang2022any, kim2022guided, kong2020diffwave, popov2021grad, miao2020flow, kim2020glow, kim2024p} and \emph{(ii)} Neural Codec-based Zero-shot TTS~\cite{wang2023neural, borsos2023soundstorm,  zhang2023speak,wang2023viola, zeghidour2021soundstream}.

Diffusion-based Zero-shot TTS models handle the problem in a non-auto regressive manner and have shown excellent performance in audio generation tasks~\cite{jeong2021diff,kulkarni2022analysis, mehta2023matcha}. Many previous works, such as~\cite{le2024voicebox, popov2021grad, liu2022diffgan}, use log Mel spectrograms as intermediate features and generate speech waveforms using high-quality vocoders. For instance, DiffVoice~\cite{liu2023diffvoice} employs a VAE-GAN autoencoder~\cite{larsen2023autoencoding} to encode the Mel-Spectrogram into a latent space, jointly modeling phoneme duration and Mel-spectrogram. FastSpeech~\cite{ren2019fastspeech} generates mel-spectrograms in parallel for faster inference, managing alignment between phoneme sequence and generated spectrogram with an explicit length regulator and duration predictor model. Flow matching training is a method that is closely related to Diffusion models offering simpler trajectories and requiring fewer function evaluations during inference~\cite{lipman2022flow, mehta2024matcha, kim2020glow}. Flow Matching (FM)~\cite{lipman2022flow} is a simulation-free approach for training continuous normalizing flows (CNFs) at scale based on regressing vector fields of fixed conditional probability paths. The relationship between the vector field and the flow $\phi$ is defined via an ordinary differential equation (ODE) $d\phi_t(y) = v_t(\phi_t(y))dt, \phi_0(y) = y$~\cite{lipman2022flow}. Benefiting from flow matching models, text-to-speech models, such as VoiceFlow~\cite{guo2023voiceflow}, MatchaTTS~\cite{mehta2023matcha} and Voicebox ~\cite{le2023voicebox}, can generate high-quality speech efficiently.

On the other side, Neural Codec-based methods formulate the TTS problem as a token-based language modeling task~\cite{wang2023neural, defossez2022high}. VALL-E~\cite{wang2023neural}, a zero-shot text-to-speech model, is a text conditioned language model trained on EnCodec tokens~\cite{defossez2022high}.  SPEAR-TTS~\cite{kharitonov2023speak} is similar to AudioLM~\cite{borsos2023audiolm} which carries out the speech generation process in two steps: first, it maps the text into discrete semantic tokens, then, in the second step, it converts the semantic tokens into acoustic tokens. In the recently proposed BASE-TTS~\cite{lajszczak2024base}, the authors propose to model the joint distribution of text tokens and discrete speech representations referred to as speechcodecs followed by a convolution-based decoder which converts these speechcodes into waveforms in an incremental, streamable manner. NaturalSpeech3~\cite{ju2024naturalspeech} combines codecs and diffusion modeling, achieving significant improvements in speech quality, prosody, intelligibility, and scalability with a 1B-parameter model trained on 200K hours of data. It decomposes speech waveforms into content, prosody, timbre, and acoustic details, reconstructing speech from these disentangled representations using a factorized diffusion model.

\subsection{Dialogue Generation}

dGSLM~\cite{nguyen2023generative} represents the pioneering textless model for generating naturalistic spoken dialogues. It utilizes a dual-tower transformer with cross-attention as its architectural backbone and leverages HuBERT ~\cite{hsu2021hubert} semantic token sequence as its input for the speech continuation task. This model generates two-channel spoken dialogue auto-regressively, without reliance on text or labels. However, its textless nature constrains its ability to direct the content of the speech it produces, occasionally leading to less intelligible outputs.

CHATS~\cite{mitsui2023towards}, while based on the same architectural principles as dGSLM, is designed to convert written dialogues into spoken conversations. It is capable of generating speech for both speaker and listener sides, conditioning on speaker ID, phoneme sequence, and context from the speaker's side, without requiring transcriptions for spontaneous behaviors or laughter. However, it does not support the capabilities of zero-shot voice cloning, relying solely on speaker ID for retaining speaker characteristics. 

SoundStorm~\cite{borsos2023soundstorm}, on the other side, is an iterative generative method that converts semantic tokens into acoustic audio tokens. It can perform zero-shot monologue and dialogue synthesis. Yet, the synthesized dialogue is generated in a sequential manner and thus sounds less realistic, lacking any spontaneous behaviors or instances of overlapping speech.




\section{CoVoMix}


Zero-shot speech generation is a task where a model synthesizes speech in a target voice that was not present in its training data. This task requires only a transcript of what is to be spoken and a speech prompt—a brief sample recording of the target voice. It is generally achieved by in-context learning with a dataset of transcribed speech $\{x, y\}$ where ${y}$ and ${x}$ denote speech utterances and their transcriptions, respectively. Zero-shot multi-talker conversational speech synthesis is designed to generate the voices of multiple speakers simultaneously, based on their transcriptions and prompts. Our approach differs from the traditional method in which  each voice is synthesized individually, and then concatenated to form a dialogue. Our goal in this work is to capture the dynamic nature of real conversations, where participants may speak over each other or respond spontaneously with interjections such as laughter.

Our proposed CoVoMix, shown in Figure~\ref{fig:intro},  consists of a multi-stream text-to-semantic model, an acoustic model and a vocoder. The text-to-semantic model first generates multi-stream semantic token sequences for each speaker, given the dialogue transcription. Then the acoustic model transforms these semantic sequences into a mixed mel-spectrogram. A vanilla HiFi-GAN vocoder~\cite{kong2020hifi} finally synthesizes mono-channel multi-round dialogue from the mel-spectrogram. We utilize a conversational dataset  $D = \{x,y\}$ for training, where $y = [y^1, y^2]$ represents a stereo dialogue featuring two speakers, and $x$  corresponds to the text transcription annotated with speaker tags.

\subsection{Multi-stream Text-to-Semantic Model}
\label{para:t2s}

The multi-stream text-to-semantic model is a sequence-to-sequence model based on encoder-decoder architecture. It takes in a text token sequence generated by a BERT text tokenizer ~\cite{devlin2018bert}, augmented with special tokens denoting speaker transitions and interjections. The output comprises a multi-stream semantic token sequence. For this study, we focus on a dual-stream setup for a dialogue between two speakers. We employ a pre-trained HuBERT model~\footnote{https://github.com/facebookresearch/fairseq/tree/main/examples/textless\_nlp/dgslm/hubert\_fisher (MIT License)}~\cite{nguyen2023generative} as a speech tokenizer to  extract the clustered discrete HuBERT hidden units as semantic token sequences and process two channels of waveform, separately. If the dialogues are captured in a single-channel recording, it is necessary to perform speaker separation to produce a dual-channel waveform in our approach. The process of semantic token extraction operates at the frame level, with a time shift of 20 milliseconds, resulting in the presence of duplicated tokens within the sequence.  We train this model on a paired speech-text dataset with cross-entropy loss, as
\begin{equation}
    \mathcal{L}_{t2s} = \sum_{c = 1}^{C} \sum_{i} \log p(s_i^{(c)} | s_{1:i-1}^{(c)} ; \theta, x)
    \label{eq:t2s}
\end{equation}
where $s_i$ is the $i$th semantic token and $c$ denotes the $c$th speaker. In order to predict two-stream semantic token sequences, we adopt a strategy wherein we divide the semantic embedding into two distinct segments (splitting it into two halves along the feature dimension) in the final linear layer of the decoder. Each segment corresponds to a different speaker participating in the conversation. This approach enables the model to capture contextual information not only from each individual speaker but also from their interaction. The dynamic exchange between speakers significantly shapes the semantic content, especially in scenarios involving multi-round conversations. 

\subsection{Acoustic Model}
\label{para:acous}
The acoustic model is a flow-matching based transformer encoder, which generates a mixed mel-spectrogram, given multi-stream semantic token sequences and multi-speaker prompts. 

At each timestamp $t \in [0,1]$, a lookup table first embeds the semantic token sequence $s = [s^1, s^2]$ into $s_{emb} =[s_{emb}^1,s_{emb}^2]$ for two speakers. We extract the corresponding mixed mel-spectrogram~$m$ and individual mel-spectrogram $[m^1,m^2]$ for each speaker of dialogue $y$. We randomly choose a mask. The masked part $\tilde{m} = m\odot mask$ is to be predicted, while the seen part $m_{ctx} = [m^1 \odot (1-mask),m^2 \odot (1-mask)]$ is considered as prompt.

At each flow step $t$, we sample $w = (1 - (1-\sigma_{min})t)\tilde{m_0} + tm$, where $\sigma_{min}$ is a hyper-parameter to control deviation and $\tilde{m_0}$ is sampled from $\mathcal{N}(m|0,\mathbf{I})$. Then, the sample $w$ at flow step $t$, the acoustic prompt $m_{ctx}$, and semantic embedding sequences $s_{emb}$ are concatenated frame-by-frame to obtain an input matrix $W_{input}$. 
Conditional Flow Matching (CFM)~\cite{lipman2022flow} is a per-example training objective, which provides equivalent gradients and does not require explicit knowledge of the intractable target vector field. Therefore, we train the acoustic model to learn the mixed mel-spectrogram with objective as in Eq.\ref{eq:acous},  where $v_t(w, m_{ctx}, s_{emb} ; \theta)$ is the transformer output with flow $w$ at step $t$.  
\begin{equation}
    \mathcal{L}_{CFM} = \mathbb{E}_{t, q(m,s),p_0(m_0)} \Vert  mask \odot ( (m - (1 - \sigma_{min})\tilde{m_0}) - v_t(w, m_{ctx}, s_{emb} ; \theta) )\Vert^2
    \label{eq:acous}
\end{equation}

During inference, to sample mixed mel-spectrogram $m$ from learned distribution $p_1(m|s,m_{ctx})$, we sample a gaussian noise $m_0$ from $p_0=\mathcal{N}(m|0,\mathbf{I})$ use an ODE solver to evaluate the flow $\phi_1(m_0)$ given $d\phi_t(m_0)/dt = v_t(w, m_{ctx}, s_{emb} ; \theta)$ and $\phi_0(m_0) = m_0$. 

We also use classifier-free guidance, a method to trade off mode coverage and sample fidelity \cite{le2023voicebox,ho2022classifier},  in the training for flow-matching model. During training, the acoustic prompt $m_{ctx}$ and semantic sequences $s_{emb}$ are dropped with $p_{uncond}$. During inference, we use the modified vector field $\tilde{v}_t(w, m_{ctx}, s_{emb}; \theta)$  shown in Equation \ref{eq:cfg} to replace $v_t(w, m_{ctx}, s_{emb}; \theta)$, where $\alpha$ is a hyperparameter controlling the strength of guidance. 
\begin{equation}
    \tilde{v}_t(w, m_{ctx}, s_{emb} ; \theta) = (1 + \alpha) v_t(w, m_{ctx}, s_{emb} ; \theta) - \alpha \tilde{v}_t(w; \theta)
    \label{eq:cfg}
\end{equation}



\section{Experimental Setup}

\subsection{Data Preparation}
\label{para:dialogue_data}
The dataset used in this work is Fisher dataset \cite{cieri2004fisher}, which is a telephone conversation dataset with 2,000h English conversations about various topics. Each dialogue was recorded in two channels with an 8kHz sample rate and an average duration of 10 minutes. We randomly divide the Fisher dataset into train/valid/test sets with 97/1/2 split. Each set has different speakers. 

\begin{figure}[htb]
  \centering
\centerline{\includegraphics[width=0.9\linewidth]{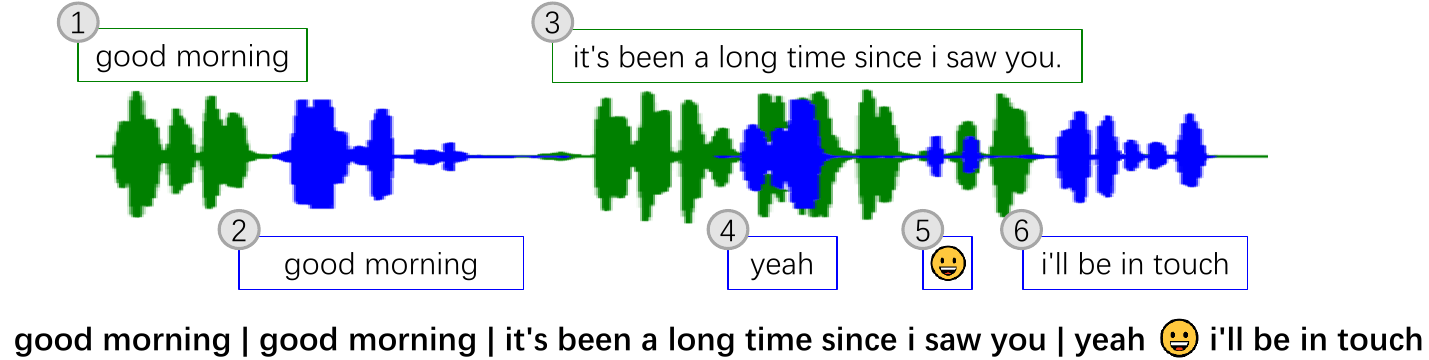}}
\caption{Dialogue transcription preparation. To better demonstrate our method, we use | and emoji to represent [spkchange] and [laughter] tokens.}
\label{fig:dataprep}
\end{figure}

The data preparation is different for monologue and dialogue. For monologue, following Nemo~\cite{kuchaiev2019nemo} script,\footnote{https://gitlab.nrp-nautilus.io/ar-noc/nemo/-/blob/master/scripts/process\_fisher\_data.py (Apache License 2.0)} we slice long dialogues into smaller mono-channel samples and concatenate them to meet the minimum duration requirement, which is set to 10 seconds by default. We prepare the corresponding transcripts, extract the mel-spectrogram and semantic token sequence for each sample. Spontaneous behavior such as laughter is labeled by [laughter] token in the transcription. For dialogue, we slice long dialogues into shorter, stereo-channel dialogues containing at least two utterances from distinct speakers. We ensure that the first and last sentences of each processed dialogue do not overlap with other dialogues, thus avoiding any extraneous content in the transcriptions. Motivated by serialized output training in speech recognition task~\cite{kanda2020serialized, li23o_interspeech}, we organize the multi-round dialogue transcript chronologically by the start time of each utterance. Two neighboring same-speaker utterances are concatenated directly, while different speakers’ utterances are separated by [spkchange] token, without explicit processing for overlap labelling. A dialogue transcription preparation example is shown in Figure \ref{fig:dataprep}. The HuBERT speech tokenizer is employed to extract the semantic tokens for each channel. Additionally, we mix audio with two channels and extract the mel-spectrogram from the mixed waveform. The detailed algorithm for dialogue data preparation is described in Appendix~\ref{appn:data}.

\subsection{Model Configurations}
\label{model_config}
We develop two text-to-semantic models, named CoSingle and CoMix,  and two acoustic models, named VoSingle and VoMix. CoSingle and VoSingle are trained exclusively on monologue data, VoMix is trained on dialogue data, and CoMix is trained on a combination of monologue and dialogue data. In addition, the vanilla HiFi-GAN vocoder is trained on monologue data.

The text-to-semantic model is a transformer-based model with rotary embedding~\cite{su2024roformer}. The encoder has 4 layers and the decoder has 4 layers. We set the dimension of text encoder and CoSingle decoder to 512, and set CoMix decoder to 1024. In order to process multi-stream for multiple talkers, CoMix applies multiple heads for generating semantic token sequences. The acoustic model is based on transformer encoder with rotary embeddings~\cite{su2024roformer} and adaptive RMSNorm~\cite{zhang2019root} for time conditioning, which has 8 layers and hidden dimension of 1024. VoMix and VoSingle have the same architecture except for the first input linear layer.  More details of model architecture is demonstrated in Appendix~\ref{appn:pipeline}.

To demonstrate the performance of our methods, the baseline that we compare with is a flow-matching speech synthesis model with phoneme representation, similar to VoiceBox~\cite{le2023voicebox}\footnote{https://github.com/lucidrains/voicebox-pytorch (MIT License)}. The baseline contains two models: the acoustic model and the duration model. The acoustic model of the baseline is the same as VoSingle model, but generates mel-spectrogram from the phoneme sequence. The duration model of baseline is to predict the duration of each phoneme, which is also trained with flow matching objective and has the same architecture with 2 layers and hidden size of 1024.

We train all models from scratch and perform inference on the best performed model on validation set. We use 8 NVIDIA TESLA V100 32GB GPUs for training. The text-to-semantic model is trained for 10 epochs with batch size 48. The acoustic model and duration model are trained for 100 epochs with batch size 64. We adopt Adam optimizer with 1e-4 learning rate. The probability of dropping condition during training is $p_{uncond} = 0.3$, and the strength of guidance is $\alpha = 0.7$ during inference.  

\subsection{System Configuration and Evaluation Setting}
\label{sec:system}

We built two systems: CoVoSingle and CoVoMix, and evaluated them on both monologue and dialogue testing sets. CoVoSingle contains CoSingle and VoSingle models.  CoVoMix system contains CoMix and VoMix. For monologue generation, CoVoSingle and CoVoMix systems directly feed the output of text-to-semantic model into the acoustic model. 
The acoustic prompt is extracted from another utterance of the target speaker. For dialogue generation, CoVoSingle generate each utterance of the dialogue and concatenate these waveforms according to the order of transcript. 
CoVoMix receives dialogue transcription as input and synthesizes mono-channel dialogue directly. The acoustic prompts are extracted from another dialogue of target speakers.

\subsection{Evaluation Metrics}
\textbf{Objective Metrics:} We use cosine speaker similarity (SIM), word error rate (WER), Mel cepstral distortion (MCD),\footnote{https://github.com/chenqi008/pymcd (MIT License)} and NISQA\footnote{https://github.com/gabrielmittag/NISQA (MIT License)} to evaluate generation results~\cite{mittag2021deep}. SIM measures the cosine similarity between speaker embeddings of generated utterance and the acoustic prompt, extracted from WavLM-TDNN~\cite{chen2022wavlm}. 
We use a market-leading Speech Recognition API for WER calculation, which measures the correctness and intelligibility. 
We use an improved MCD metric that adopts the Dynamic Time Warping (DTW) algorithm to find the minimum MCD between two speeches~\cite{battenberg2020location}. 
NISQA~\cite{mittag2021deep} measures the speech quality and naturalness of the synthesized speech. 

\textbf{Subjective Metrics: } We perform a human evaluation on the generated monologue and dialogue examples. For monologue, we measure naturalness using comparative mean option score (CMOS).  For dialogue, we use CMOS to measure naturalness and how seamlessly the conversation flows. We use the similarity mean option score (SMOS) between the synthesized and prompt speech to measure the speaker similarity for both monologue and dialogue. 14 professional linguistic experts provide judges for all subjective evaluations. They provide a rating to the second audio, which is randomly selected from a pair of audios, in the (-3 to +3) range. The instructions of subjective evaluations are provided in Appendix \ref{appn:evaluation}.

\textbf{Dialogue Metrics: } We assess the naturalness of the generated dialogue speech through three metrics: 1) \textit{Turn-taking Statistics:} By employing a pre-trained speaker diarization model ~\cite{plaquet2023powerset,bredin2023pyannote},\footnote{https://github.com/pyannote/pyannote-audio (MIT License)} we measure the duration of inter- and intra-speaker silences, overlapped speech, and active speech.
2) \textit{Para-linguistic Behaviors:} Our evaluation focuses on laughter in this study. Employing a laughter detection tool \cite{gillick2021robust},\footnote{https://github.com/jrgillick/laughter-detection (MIT License)} we identify instances of laughter and calculate both the total count and average duration of these events.
and 3) \textit{Speech Consistency:} To evaluate consistency, we generate ten dialogues, each containing more than five utterances from the target speaker. We then select five three-second segments at random from the target speaker and compare the cosine similarity of speaker embeddings among these segments.

\section{Result and Analysis}
 
\subsection{Objective and Subjective Metrics}
Table \ref{tab:main} shows objective and subjective evaluation results for monologue and dialogue generation across various systems.

We observe that the systems leveraging our proposed methods, i.e., CoVoSingle and CoVoMix, achieve higher speaker similarity, lower WER and MCD than baseline on monologue evaluation set. The phoneme-based baseline model requires accurate phoneme-level alignment, however, it is challenging to perform accurate forced-alignment using conventional alignment tool~\cite{mcauliffe2017montreal},\footnote{https://github.com/MontrealCorpusTools/Montreal-Forced-Aligner (MIT License)}  especially for speech with spontaneous behavior and noisy background. These inaccuracies in alignment can lead to significant performance degradation. By substituting phoneme representation with semantic token sequences, our approach eliminates the dependency on phoneme-level alignment, thereby enhancing model performance.  

The dialogue results show that, unlike monologue, the ground truth and CoVoMix exhibit high WER due to overlapping speech segments. The transcriptions are chronologically sorted, leading to mismatches between transcription and speech in overlapped parts. Furthermore, automatic recognizing overlapped speech while maintaining a low WER remains a challenging task to date. CoVoSingle, which generates utterances separately and combines them, avoids this issue, resulting in lower WER.
 
In terms of speech quality, we observe that the proposed systems can surpass the ground truth on both monologue and dialogue sets. The flow-matching based acoustic model is able to eliminate background noise, and therefore produces cleaner audio than real data. CoVoMix can generate overlapped speech, which may result in a slightly lower NISQA, comparing with CoVoSingle.

\begin{table}[!ht]
    \centering
    \caption{Objective and subjective evaluation results for monologue and dialogue generation across various systems.The symbol "{$\dagger$}" is used to indicate that the system performance is significantly different (p<0.01) from CoVoSingle system in terms of CMOS and SMOS scores. }
    \label{tab:main}
    \begin{tabular}{c|c|cccc|cc}
    \toprule
        Eval Set & System & SIM $\uparrow$ & WER $\downarrow$ & MCD $\downarrow$ & NISQA $\uparrow$ & CMOS $\uparrow$ & SMOS $\uparrow$ \\ \midrule
        \multirow{4}{*}{Monologue} & GroundTruth & 0.59  & 6.10 & / & 3.03 & / & / \\ 
        ~ & Baseline & 0.42  & 15.85 & 9.45 & 2.93    & -1.60$\dagger$ & -1.18$\dagger$\\ 
        ~ & CoVoSingle & 0.49  & 9.99  & 6.15 & 3.04   & 0.00 & 0.00 \\ 
        ~ & CoVoMix & 0.49  & 8.95  & 6.04 & 3.01  & 0.83$\dagger$ & 0.11 \\
        \midrule
        \multirow{3}{*}{Dialogue} & GroundTruth  &  / & 14.91 & / & 2.73  & /& /    \\ 
        ~ & CoVoSingle & / & 11.77  & 6.91 & 2.90  & 0.00 & 0.00\\ 
        ~ & CoVoMix & / & 19.84  & 6.82 & 2.87   &  0.81$\dagger$ & 0.60$\dagger$\\ 
        \bottomrule
    \end{tabular}
\end{table}

The subjective evaluations consistently support the findings of the objective metrics. As shown in Table \ref{tab:main}, CoVoSingle significantly outperforms baseline in terms of both CMOS and SMOS scores for monologue testing set. Furthermore, across both monologue and dialogue testing sets, CoVoMix demonstrates significantly better performance over CoVoSingle. 

We have not found a good way to measure the objective similarity metric for the dialogue testing set due to the necessity of speaker diarization, since the potential errors in speaker diarization could impact the fairness of the comparison. Therefore, for the dialogue SMOS evaluation, testing dialogues were manually segmented into multiple single-speaker utterances to avoid speaker diarization errors.


\subsection{Dialogue Metrics}
\subsubsection{Turn-taking Statistics}
We define four turn-taking activities in a dialogue: 1) intra speaker pause (silence between active speech of the same speaker), 2) inter speaker silence (silence between active speech of different speaker), 3) overlapped segments, and 4) active speech of each speaker~\cite{heldner2010pauses, nguyen2023generative}. 

\begin{figure}[htb]
  \centering
\centerline{\includegraphics[width=1.0\linewidth]{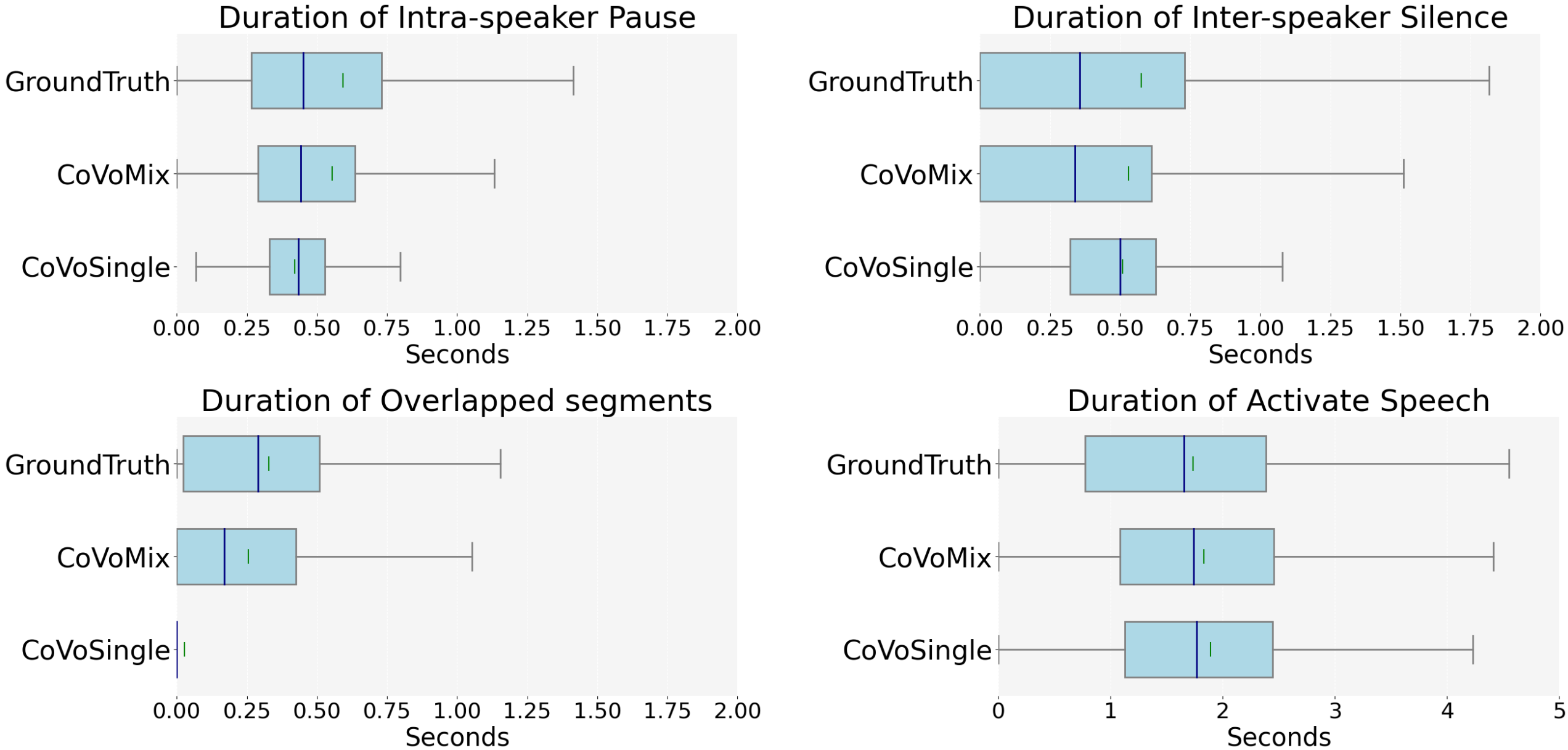}}
\caption{Distribution of durations of turn-taking events across models. The blue line and the green line represent the median and mean of each event. The more similar to groundtruth, the better. }
\label{fig:statistic}
\end{figure}

Figure \ref{fig:statistic} shows the distribution of various turn-taking activities. The degree of similarity to the ground truth reflects the model's ability to simulate turn-taking in a dialogue. While CoVoSingle can synthesize high-quality monologue, it exhibits subpar performance in dialogue turn taking events, particularly in managing intra-speaker pause, inter-speaker silence, and overlap control. Simply concatenating monologue at utterance level results in low variance in inter- and intra- speaker silence distribution, leading in a dialogue which sounds robotic and lacks the natural flow of conversation~\cite{nguyen2023generative}. In contrast, CoVoMix demonstrates a high similarity to the ground truth in these turn-taking events, yielding more human-realistic dialogues.


\subsubsection{Para-linguistic Behaviors}
\begin{figure}[htb]
  \centering
\centerline{\includegraphics[width=0.9\linewidth]{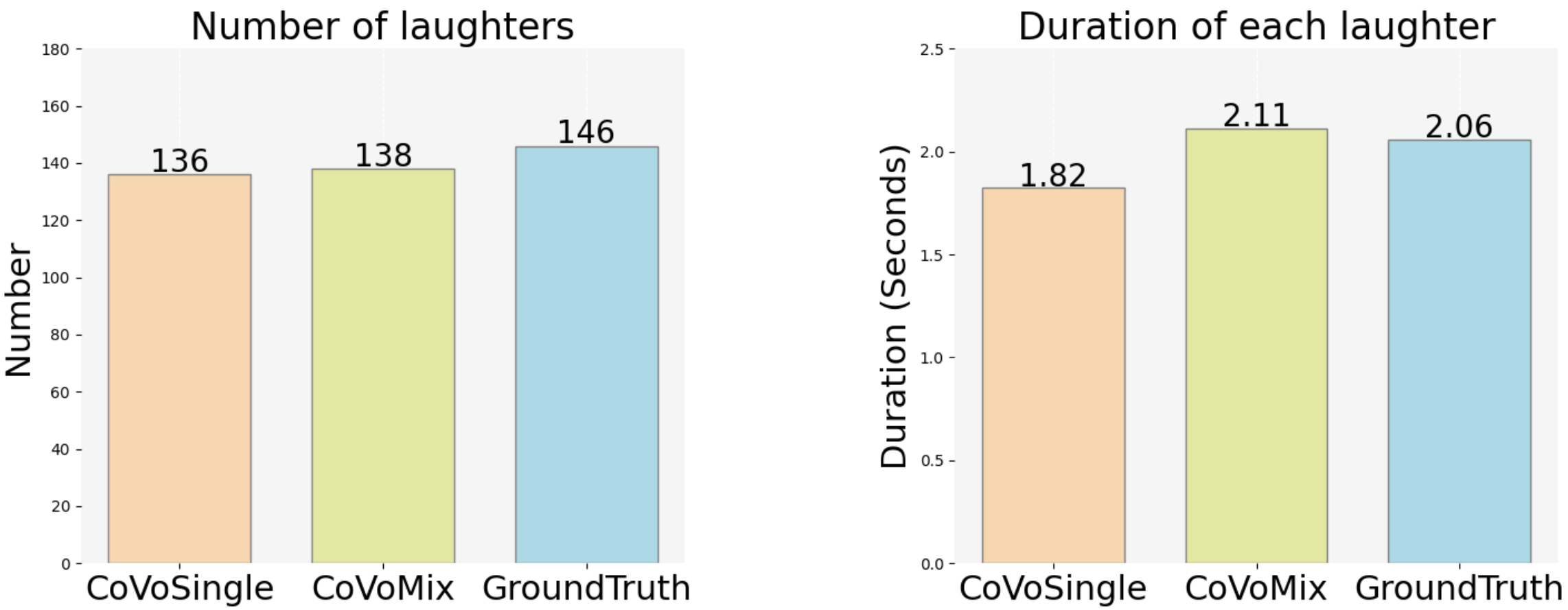}}
\caption{Comparison of number and duration of laughter among models}
\label{fig:laughter}
\end{figure}

We computed the frequency and duration of spontaneous laughter behaviors across the conversation test set and compared these metrics across models to check their closeness to the ground truth. As illustrated in Figure \ref{fig:laughter}, it shows that all proposed models can generate laughter with a frequency close to the ground truth, demonstrating precise control over these human-like behaviors. Moreover, CoVoMix can produce dialogues with an average laughter duration that is closer to the ground truth, whereas CoVoSingle tends to synthesize shorter instances of laughter.

\subsubsection{Speech Consistency}
\begin{figure}[htb]
  \centering
\centerline{\includegraphics[width=0.85\linewidth]{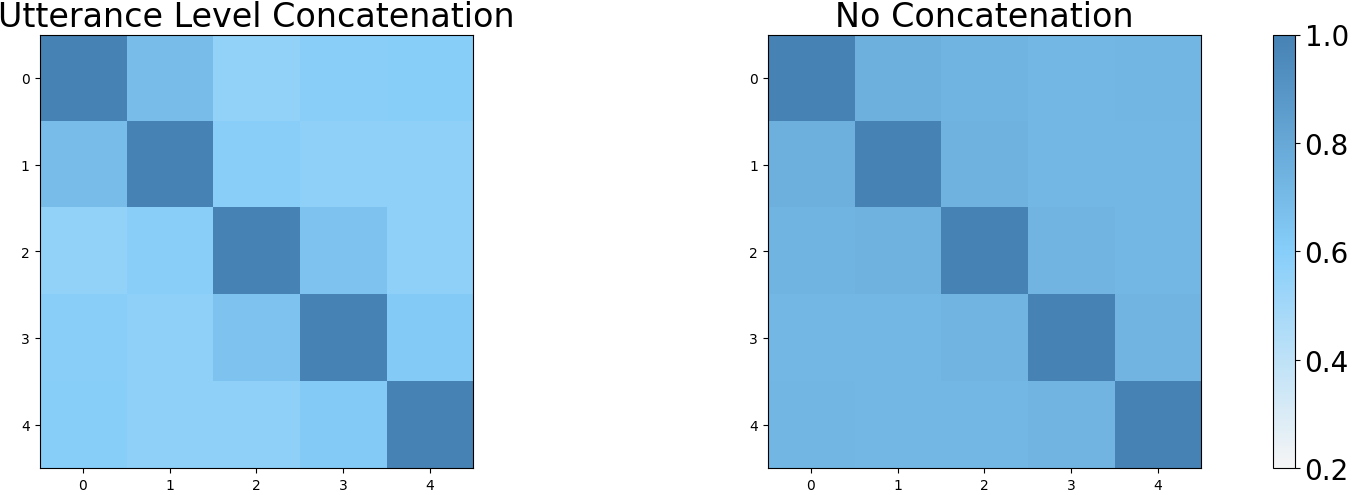}}
\caption{Speech consistency of CoVoSingle and CoVoMix for dialogue generation}
\label{fig:consistency}
\end{figure}

We calculate the speaker similarity between any two pairs of different utterances in a long conversation. Figure \ref{fig:consistency} presents a heatmap of the cosine similarity between different segments, contrasting utterance-level concatenation methods like CoVoSingle with non-concatenation approaches like CoVoMix. A lighter shade indicates lower speaker similarity. The figure's color inconsistencies reveal that utterance-level concatenation can indeed lead to dissimilar speaker characteristics, particularly for non-adjacent utterances. Generating the entire dialogue without concatenation results in significantly improved consistency of speaker similarity across various utterances.

\section{Ablation Studies and Extension}
To enhance the effectiveness of text-to-semantic modeling, we conducted ablation studies focusing on data augmentation and model size.  In addition to real dialogue data, we incorporated simulated dialogues and monologue sentences into training data. Results show the benefits of such augmentation, as evidenced by improved model prediction accuracy, i.e., reduced WER, in both monologue and dialogue generation tasks. Furthermore, we explored the impact of output channel configurations for the acoustic model by comparing single-channel mixed speech output with dual-channel outputs, where each channel contained speech from an individual speaker. Experimental results show that dual-channel outputs underperformed in WER, and outperformed in NISQA compared with single-channel outputs. Please refer to Appendix \ref{appn:ablation} for the detailed results of all ablation studies.

Our acoustic model can generate specific speakers' voices, given semantic token sequences and target speakers' prompts. So it is straightforward to be extended to a voice conversion task, which modifies the speech of a source speaker and makes their speech sound like that of another target speaker without changing the content information. Instead of predicting semantic tokens from given text, we extract the semantic tokens from the speech of the source speaker. 
VoSingle performs voice conversion of dialogue by processing each channel individually and then mix them up, while VoMix model achieves voice conversion simultaneously. 
We notice that in addition to achieving high speaker similarity, these systems can also achieve high spectral similarity, indicating the strong zero-shot voice conversion capability. Moreover, VoMix performs better than VoSingle in both monologue and dialogue sets. The detailed results are shown in Appendix \ref {appn:vc} and the corresponding demo is provided in https://aka.ms/covomix. 


\section{Conclusion, Limitation, Future Work and Broader Impacts}
We introduce the CoVoMix system for human-like monologue and dialogue generation. The system is composed of an auto-regressive text-to-semantic model and a flow-matching based acoustic model, with semantic token sequence as an intermediate representation. 
A 2k-hour conversational telephone speech dataset is leveraged in training these two models of CoVoMix.
Through both objective and subjective evaluations, CoVoMix not only achieves high naturalness and zero-shot speaker similarity in both monologue and dialogue generations but also demonstrates its proficiency in the fluency of dialogue turn-taking and spontaneous behavior generation.

\label{para:limitation}
\textbf{Limitation and Future work} We have observed instances of words being omitted or duplicated occasionally in synthesized speech. This is primarily attributed to the text-to-semantic model being an auto-regressive model without forced duration. Additionally, the dataset utilized for this study is sampled at 8 kHz with background noise, factors that contribute to the degradation of speech quality. In future work, we aim to enhance the text-to-semantic model by scaling it up or initializing it with a pre-trained model, and employing super-resolution methods to improve the training data fidelity.

\textbf{Broader Impacts} A high-quality and human-like speech generation model like CoVoMix can enable many applications that improve the quality of our life. However, since CoVoMix could synthesize speech that maintains speaker identity, it may carry potential risks in misuse of the model, such as spoofing voice identification or impersonating a specific speaker. To mitigate such risks, it is possible to build a detection model to discriminate whether an audio clip was synthesized by CoVoMix.

\newpage
\bibliographystyle{IEEEtran}
\bibliography{mybib}

\newpage

\appendix

\section{Model Architecture}
\label{appn:pipeline}
\begin{figure}[htbp]
  \centering
\centerline{\includegraphics[width=1.0\linewidth]{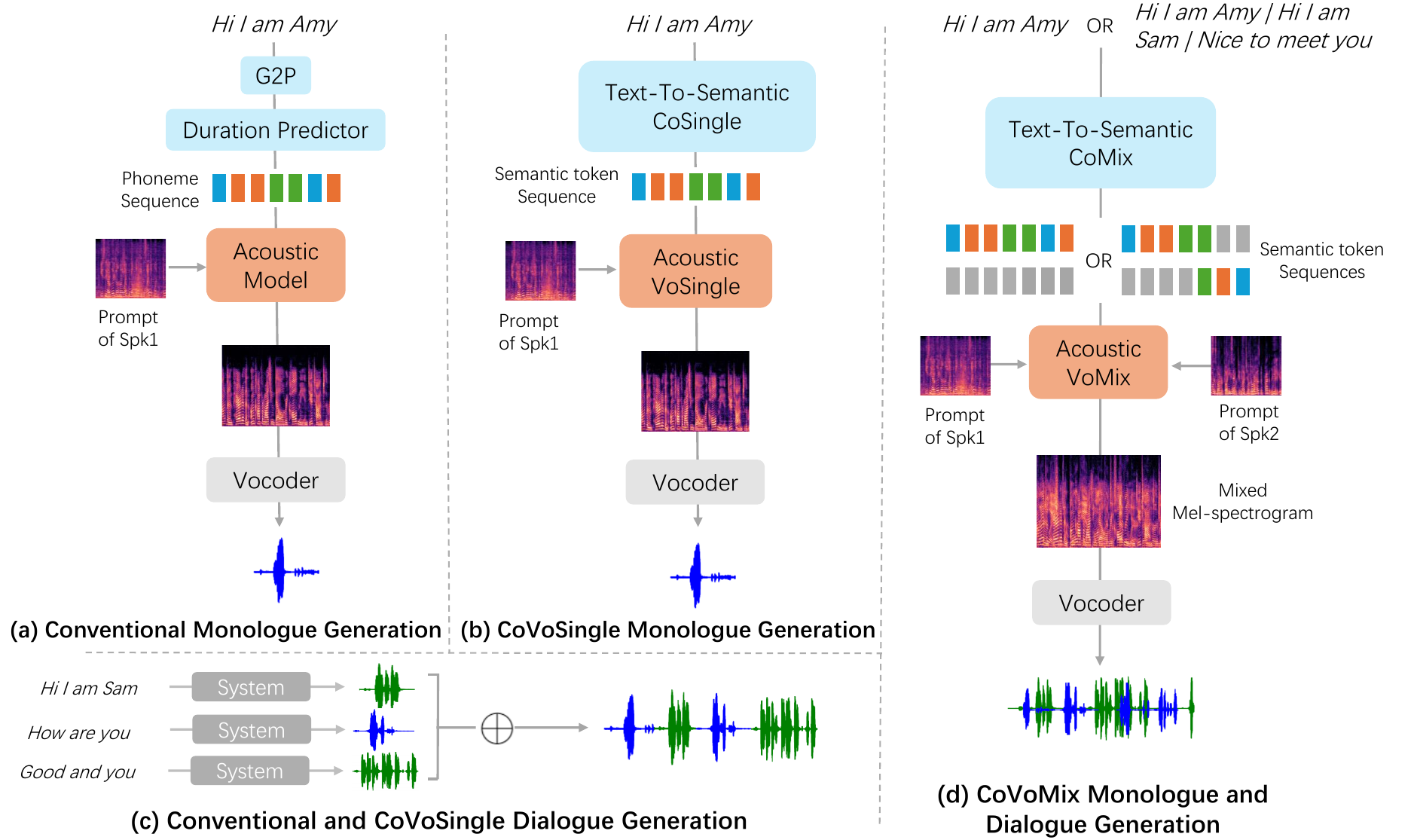}}
\caption{Comparison of generation pipeline among conventional method and our proposed CoVoSingle and CoVoMix methods.  
}
\label{fig:pipeline}
\end{figure}

Figure \ref{fig:pipeline} compares the generation pipeline among conventional method and our proposed CoVoSingle and CoVoMix methods. Figure \ref{fig:pipeline}(a) shows the conventional monologue generation process with phoneme representation. (b) shows our proposed CoVoSingle approach for monologue generation. (c) demonstrates the concatenation method for conventional and CoVoSingle models to generate dialogue. (d) shows the architecture of our proposed CoVoMix model for monologue and dialogue generation.

\begin{figure}[htb]
  \centering
\centerline{\includegraphics[width=0.9\linewidth]{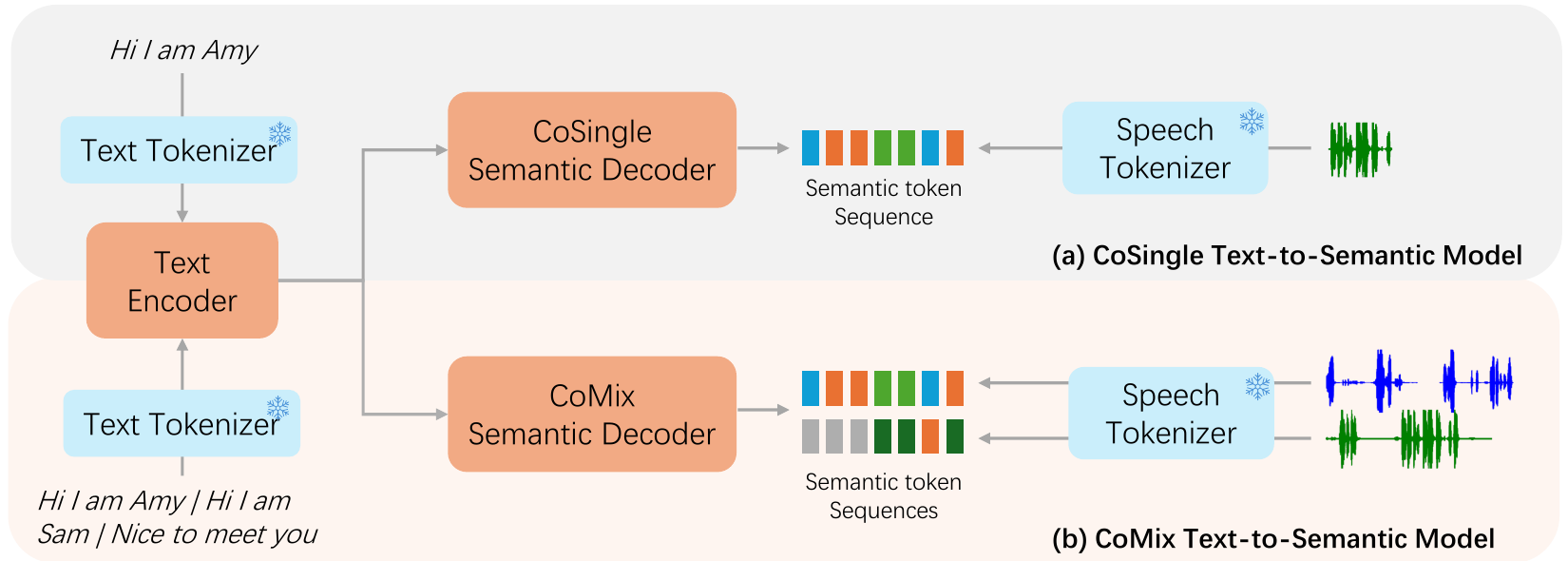}}
\caption{Text-to-semantic model}
\label{fig:t2s}
\end{figure}

Figure \ref{fig:t2s} shows the architecture of text-to-semantic model.  We propose two types of text-to-semantic model: CoSingle and CoMix. CoSingle model has single-stream decoder, while CoMix applies multi-stream decoder to generate multiple semantic token sequences for different speakers, as introduced  in Section\ref{para:t2s}. 

\begin{figure}[htb]
  \centering
\centerline{\includegraphics[width=1.0\linewidth]{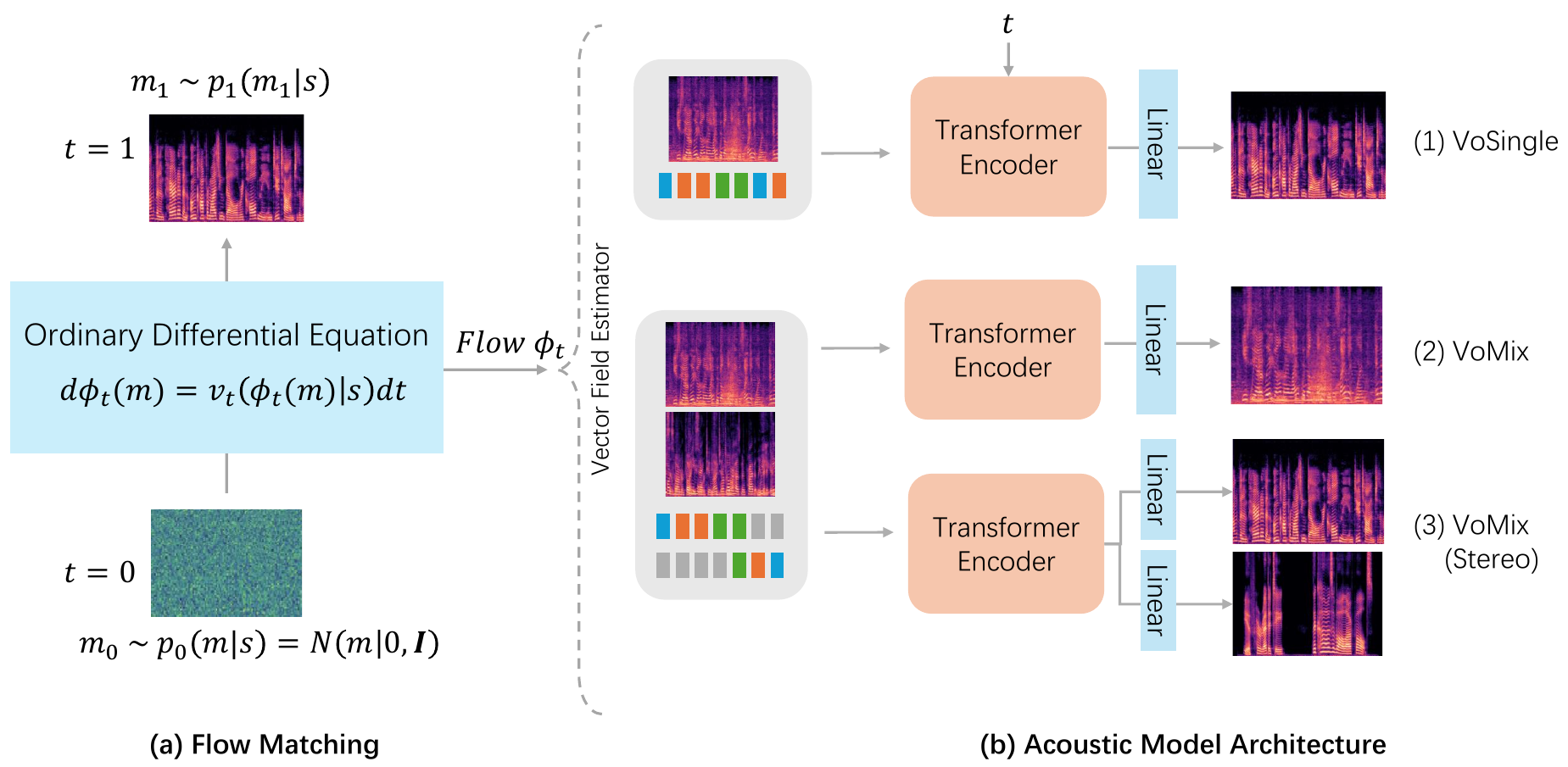}}
\caption{The acoustic model}
\label{fig:acous}
\end{figure}

Figure \ref{fig:acous} shows the architecture of acoustic model, a flow-matching based transformer encoder. We propose three types of acoustic model: VoSingle, VoMix and VoMix-stereo. VoSingle is a single stream transformer encoder to generate single talker mel-spectrogram. VoMix and VoMix-stereo have the same architecture except for the last linear layers, which generate mono channel mixed mel-spectrogram and multiple single talker  mel-spectrograms respectively.

\section{Additional Experiments}
\label{appn:ablation}
We perform detailed ablation experiments for model combination, model size and training data. 



\subsection{Ablation Study on Model Combination}
Table \ref{tab:stereo} illustrates our proposed systems in monologue and dialogue evaluation set, which are a combination of text-to-semantic model and acoustic model. We abbreviate text-to-semantic as T2S and stereo as S.

For monologue generation, CoVoSingle and CoVoMix systems directly feed the output of text-to-semantic model into the acoustic model. However, CoVoSinx model set the second semantic sequence as all silence tokens.  The acoustic prompt is extracted from another utterance of the target speaker. 

We first observe that when using the same text-to-semantic model, different acoustic models influence WER, indicating that the pronunciation and rhyme also affect the intelligibility of speech. For example, CoVoSinx achieves better speech intelligibility than CoVoSingle due to the use of VoMix. The speaker similarity performance for monologue generation is the similar across models. 

For dialogue generation, we notice that stereo systems show comparable WER than mono systems. Moreover, we observe that stereo systems show higher NISQA speech quality, indicating that predicting each channel separately causes less distortion in speech quality than predicting mixed mel-spectrogram.  


\begin{table}[!ht]
    \centering
    \caption{Objective evaluation on monologue and dialogue for mono and stereo acoustic model}
    \label{tab:stereo}
    \begin{tabular}{l|c|c|ccc|cc}
    \toprule
\multirow{2}{*}{System} & \multirow{2}{*}{T2S} & \multirow{2}{*}{Acousic} & \multicolumn{3}{c|}{Monologue} & \multicolumn{2}{c}{Dialogue} \\
        ~ & ~& ~ & SIM $\uparrow$ & WER $\downarrow$ & NISQA $\uparrow$ & WER $\downarrow$ & NISQA $\uparrow$ \\ \midrule
         GroundTruth & / & /&  0.59  & 6.10  & 3.03 &  14.91 & 2.73  \\ \midrule
         CoVoSingle & \multirow{3}{*}{CoSingle} & VoSingle &  0.49 & 9.99 & 3.01 & 11.76 & 2.90  \\
         CoVoSinx & ~ & VoMix  & 0.49  & 8.78  & 3.12 &  12.27 & 2.97  \\ 
         CoVoSinx-S  & ~ & VoMix-S  & /  & /  & / &  12.95 & 3.19  \\ \midrule
         CoVoMix & \multirow{2}{*}{CoMix} & VoMix & 0.49  & 8.95  & 3.01 & 19.84 & 2.87   \\
           CoVoMix-S& ~ & VoMix-S & /  & / & /  & 20.35 & 3.00 \\
         \bottomrule
    \end{tabular}
\end{table}


\subsection{Ablation Study on Discrete Semantic Representation and Model Size }

\begin{figure}[htb]
  \centering
\centerline{\includegraphics[width=0.8\linewidth]{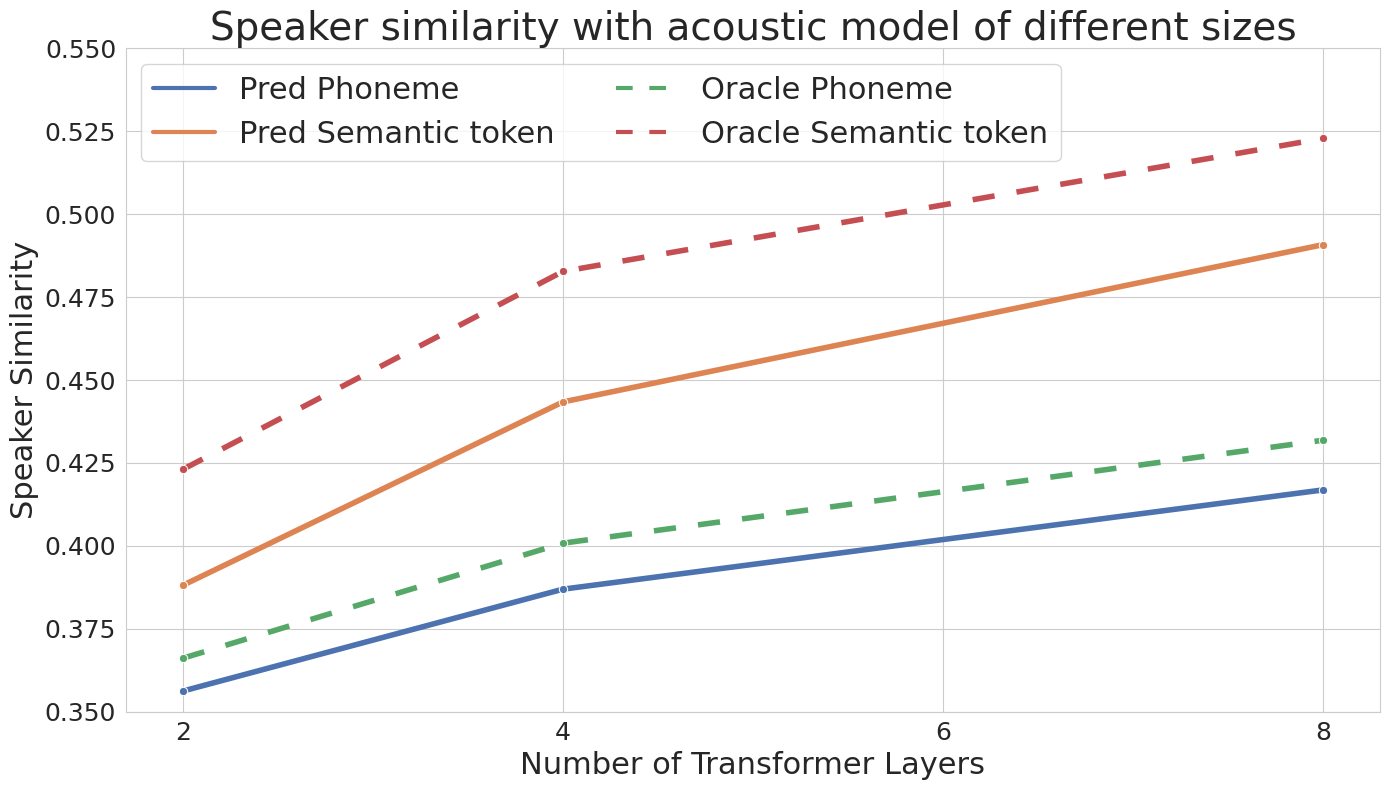}}
\caption{Speaker similarity across acoustic model of different size}
\label{fig:spksim}
\end{figure}

Figure \ref{fig:spksim} compares the speaker similarity of oracle phoneme, predicted phoneme using duration predictor, oracle semantic token, and predicted semantic token using text-to-semantic model under the same architecture of acoustic model.

First, we observe that larger acoustic model improves the speaker similarity of the generated speech as the model layer deepens. Second, the similarity using semantic token sequences is higher than using phoneme and even exceeds oracle phoneme representations. This demonstrates the advantages of using semantic token sequences, which not only avoids forced-alignment and improves the word error rate, but also improves the model's speaker modeling capabilities. Third, for both phoneme and semantic token, the predicted representations are not as good as oracle representations. The duplicated semantic token sequence is more difficult to predict, leading to a bigger gap between oracle and prediction, indicating further improvement space for the performance and accuracy of text-to-semantic model in the future.

\subsection{Ablation Study on Data Augmentation for Text-To-Semantic Model }


When the total amount of data is deterministic, the diversity of data, such as data duration and the dialogue content, becomes important. Table \ref{tab:ablation_data} shows the performance of text-to-semantic model with different data augmentation method on monologue and dialogue generation in terms of correctness. In this study, the acoustic model used only contains 2 transformer layers instead of 8 layers mentioned in \ref{model_config} for faster inference. We incorporate a diverse training dataset including both short and long sentences for the text-to-semantic model to accurately generate dialogues of varying lengths. The long monologue has a minimum duration of 10 seconds, whereas the short monologue has minimum duration of 1 second. Additionally, we enhance data variety by simulating dialogues from monologues through concatenation, which also improved the prediction accuracy of semantic token sequences. As a result of utilizing monologue data of different lengths and synthetic dialogue data, the text-to-semantic model demonstrated the best performance.

\begin{table}[!ht]
    \centering
    \caption{Ablation study of data augmentation methods}
    \setlength\tabcolsep{4pt}\label{tab:ablation_data}
    \begin{tabular}{cccc|cc}
    \toprule
 \multicolumn{4}{c|}{Training Data} & \multicolumn{2}{c}{WER  $\downarrow$ }  \\ 
       Real Dialogue & Simu Dialogue & Short Monologue & Long Monologue  & Monologue & Dialogue  \\ \midrule
        \checkmark & $\times$ & $\times$ & $\times$ & 28.50 & 22.30  \\ 
         \checkmark & $\times$ & \checkmark &  $\times$ & 11.99 & 21.82  \\ 
          \checkmark & \checkmark & \checkmark & $\times$ &   11.22 & 20.79  \\ 
          \checkmark & \checkmark & \checkmark &  \checkmark & 10.54 & 20.87  \\ 
        \bottomrule
    \end{tabular}
\end{table}





\section{Extension to Voice Conversion}
\label{appn:vc}
In addition to zero shot speech synthesis, our methods can also achieve voice conversion for single person monologues and multi talker conversations. VoSingle performs voice conversion of dialogue by processing each channel individually and then mix them up, while VoMix model achieves voice conversion simultaneously.   

Table \ref{tab:vc} demonstrates the objective results in monologue and dialogue scenario. 
We notice that in addition to achieving high speaker similarity, these systems can also achieve high spectral similarity, indicating the strong zero-shot voice conversion capability of our proposed system. Moreover, VoMix performs better than VoSingle in both monologue and dialogue sets.

\begin{table}[!ht]
    \centering
    \caption{Objective evaluation on voice conversion for monologue and dialogue generation. The symbol "{$\dagger$}" is used to indicate that the system performance is significantly different (p<0.01) from VoSingle system}
    \label{tab:vc}
    \begin{tabular}{l|ccc}
    \toprule
          Type & System & SIM $\uparrow$ & MCD $\downarrow$  \\ \midrule 
          \multirow{2}{*}{Monologue} & VoSingle & 0.47 & 6.47  \\ 
        ~ & VoMix & 0.49$\dagger$ & 6.46  \\  \midrule 
        \multirow{2}{*}{Dialogue} & VoSingle & / & 6.70  \\
        ~ & VoMix & / & 6.59$\dagger$   \\ 
         \bottomrule
    \end{tabular}
\end{table}

\section{Experiment Statistical Significance}

In order to determine  statistical significance of the main experiment in Table \ref{tab:main}, we first use \textit{numpy.mean} and \textit{numpy.std} in python to calculate the mean and the standard deviation of objective metrics across the test set in Table \ref{tab:dialogue_statistical}. Moreover, we use z-test to determine if the differences are statistically significant. We notice that results are statistically significant for WER, MCD and NISQA in both monologue and dialogue evaluation sets. Similar to subjective evaluation result, the speaker similarity performance of the CoVoSingle and CoVoMix systems are relatively close and do not show significant differences. Besides, the WER has a large deviation because the text-to-semantic model might synthesize speech with omitted or duplicated words, as mentioned in the limitation part in Section \ref{para:limitation}.

\begin{table}[!ht]
    \centering
    \caption{Objective evaluation results for monologue and dialogue generation across various systems. The symbol "{$\dagger$}" is used to indicate that the system performance is significantly different (p<0.01) from CoVoSingle system}
    \label{tab:dialogue_statistical}
    \begin{tabular}{c|c|cccc}
    \toprule
        Eval Set & System & SIM $\uparrow$ & WER $\downarrow$ & MCD $\downarrow$ & NISQA $\uparrow$  \\ \midrule
        \multirow{2}{*}{Monologue} & CoVoSingle & 0.49±0.17  & 9.99±9.02  & 6.15±1.85 & 3.04±0.39    \\ 
        ~ & CoVoMix & 0.49±0.18  & 8.95±8.68$\dagger$  & 6.04±2.03$\dagger$  & 3.01±0.44$\dagger$   \\
        \midrule
        \multirow{2}{*}{Dialogue} & CoVoSingle & / & 11.77±9.00  & 6.91±1.87 & 2.90±0.28  \\ 
        ~ & CoVoMix & / & 19.84±19.83$\dagger$   & 6.82±2.12$\dagger$   & 2.87±0.37$\dagger$    \\ 
        \bottomrule
    \end{tabular}
\end{table}

To investigate the effect of randomness, we assess the same model using three different random seeds. For each seed, we evaluate the model performance and compute the mean and standard deviation across different seeds. The results are shown in Table \ref{tab:var_seed}. Our findings indicate that the standard deviation among the different random seeds is relatively small, suggesting that the system exhibits stability in the presence of randomness.

\begin{table}[!ht]
    \centering
    \caption{Objective evaluation results for monologue and dialogue generation across various systems with different random seeds}
    \label{tab:var_seed}
    \begin{tabular}{c|c|cccc}
    \toprule
        Eval Set & System & SIM $\uparrow$ & WER $\downarrow$ & MCD $\downarrow$ & NISQA $\uparrow$  \\ \midrule
        \multirow{2}{*}{Monologue} & CoVoSingle & 0.484±0.005  & 10.206±0.166  & 6.147±0.017 & 3.035±0.014    \\ 
        ~ & CoVoMix & 0.488±0.005  & 9.378±0.349  & 6.058±0.015 & 3.008±0.002   \\
        \midrule
        \multirow{2}{*}{Dialogue} & CoVoSingle & / & 11.903±0.262  & 6.916±0.031 & 2.902±0.002  \\ 
        ~ & CoVoMix & / & 19.542±0.587 & 6.829±0.006  & 2.870±0.008   \\ 
        \bottomrule
    \end{tabular}
\end{table}

\section{Data Preparation}
\label{appn:data}

Algorithm \ref{algo:dialoguedata} illustrates the dialogue data preparation pipeline, introduced in Section\ref{para:dialogue_data}. The hyper-parameter maxDuration is set to 40 seconds by default. 

\begin{algorithm}
\caption{Dialogue Data Preparation}
\label{algo:dialoguedata}
\begin{algorithmic}[1]
\REQUIRE Dialogue recordings $y$, corresponding transcriptions $x$, maxDuration.
\STATE Segment dialogues into utterances per speaker: $y^A$, $y^B$ with transcripts $x^A$, $x^B$ and identity $z$.
\STATE Sort $x^A, x^B, y^A, y^B$ by start times into sequences $\mathbf{X}, \mathbf{Y}, \mathbf{Z}$.
\STATE Initialize $cache = []$, $spkcache = []$, $OutputDialogue = []$, $OutputTranscript = []$.
\FOR{each $(x_{new}, y_{new}, z_{new})$ in zip($\mathbf{X}, \mathbf{Y}, \mathbf{Z}$)}
    \IF{$cache$ is empty}
        \STATE Add $(x_{new}, y_{new}, z_{new})$ to $cache$; add $z_{new}$ to $spkcache$.
    \ELSIF{StartTime$(y_{new}) >$ EndTime$(cache[-1])$ AND $|set(spkcache)| > 1$}
        \STATE Compile dialogue from $cache$.
        \STATE Compile transcription from $cache$ by start time with speaker change symbol.
        \STATE Reset $cache$ and $spkcache$.
        \STATE Add compiled dialogue to $OutputDialogue$, transcription to $OutputTranscript$.
    \ELSIF{EndTime$(cache[-1]) - $ StartTime$(cache[0]) >$ maxDuration}
        \STATE Reset $cache$ and $spkcache$.
    \ELSE
        \STATE Continue populating $cache$ and $spkcache$.
    \ENDIF
\ENDFOR
\RETURN $OutputDialogue$, $OutputTranscript$.
\end{algorithmic}
\end{algorithm}



\section{Subjective Evaluation Instruction}
\label{appn:evaluation}
\begin{figure}[htb]
  \centering
\centerline{\includegraphics[width=0.95\linewidth]{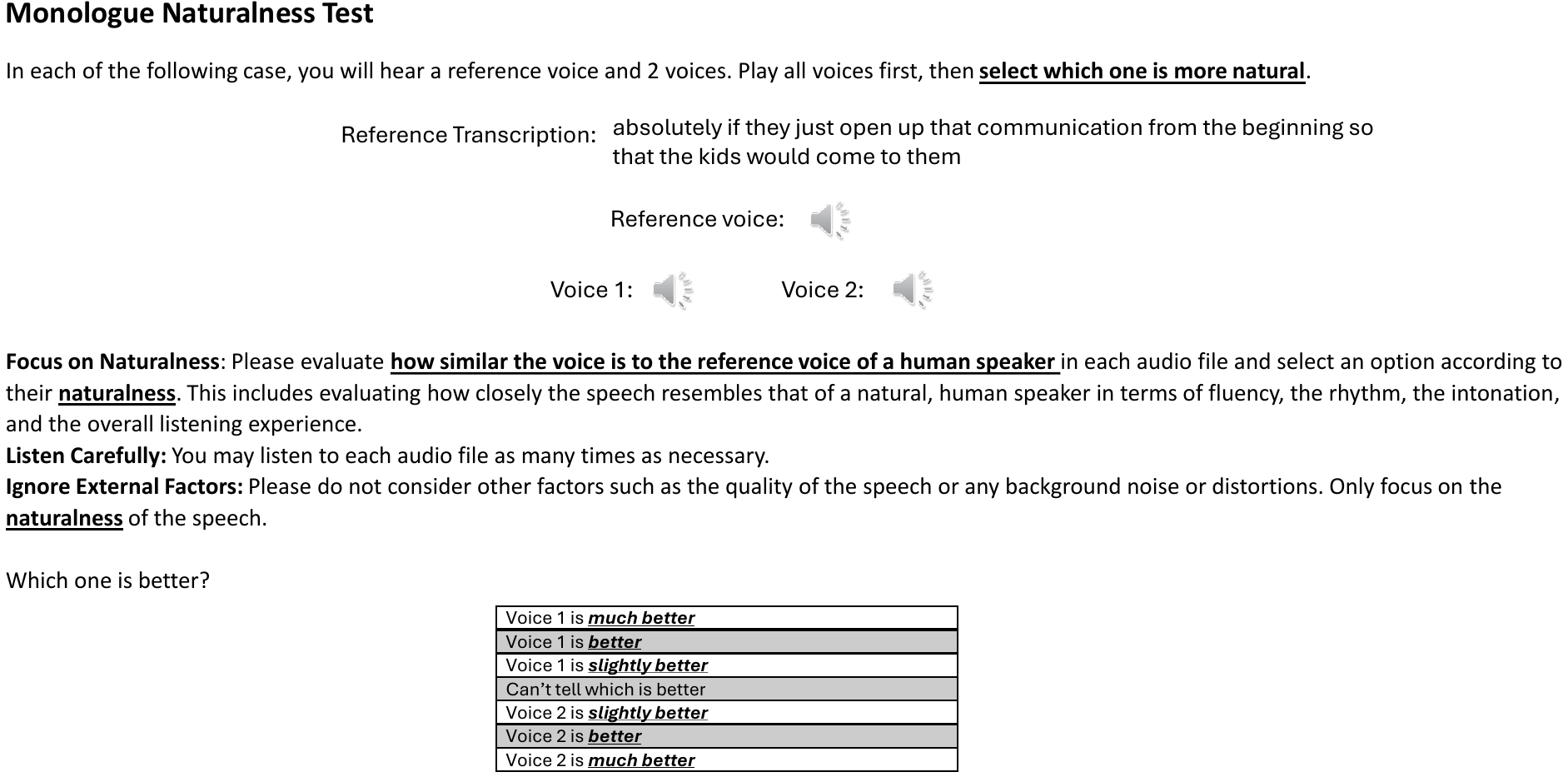}}
\caption{CMOS evaluation   template for monologue generation}
\label{fig:mono-cmos}
\end{figure}

\begin{figure}[H]
  \centering
\centerline{\includegraphics[width=0.95\linewidth]{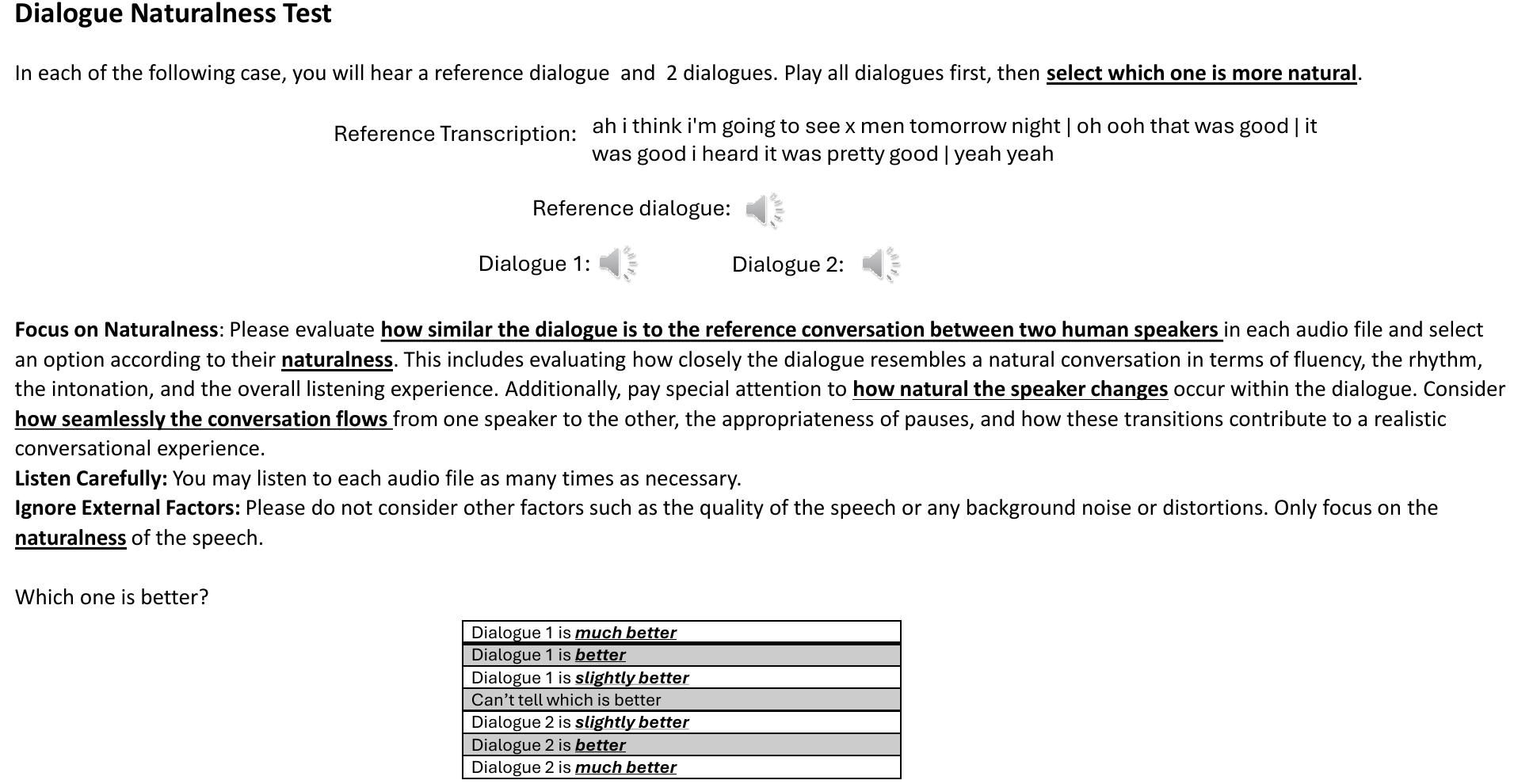}}
\caption{CMOS evaluation template for dialogue generation}
\label{fig:dia-cmos}
\end{figure}

\begin{figure}[htb]
  \centering
\centerline{\includegraphics[width=0.95\linewidth]{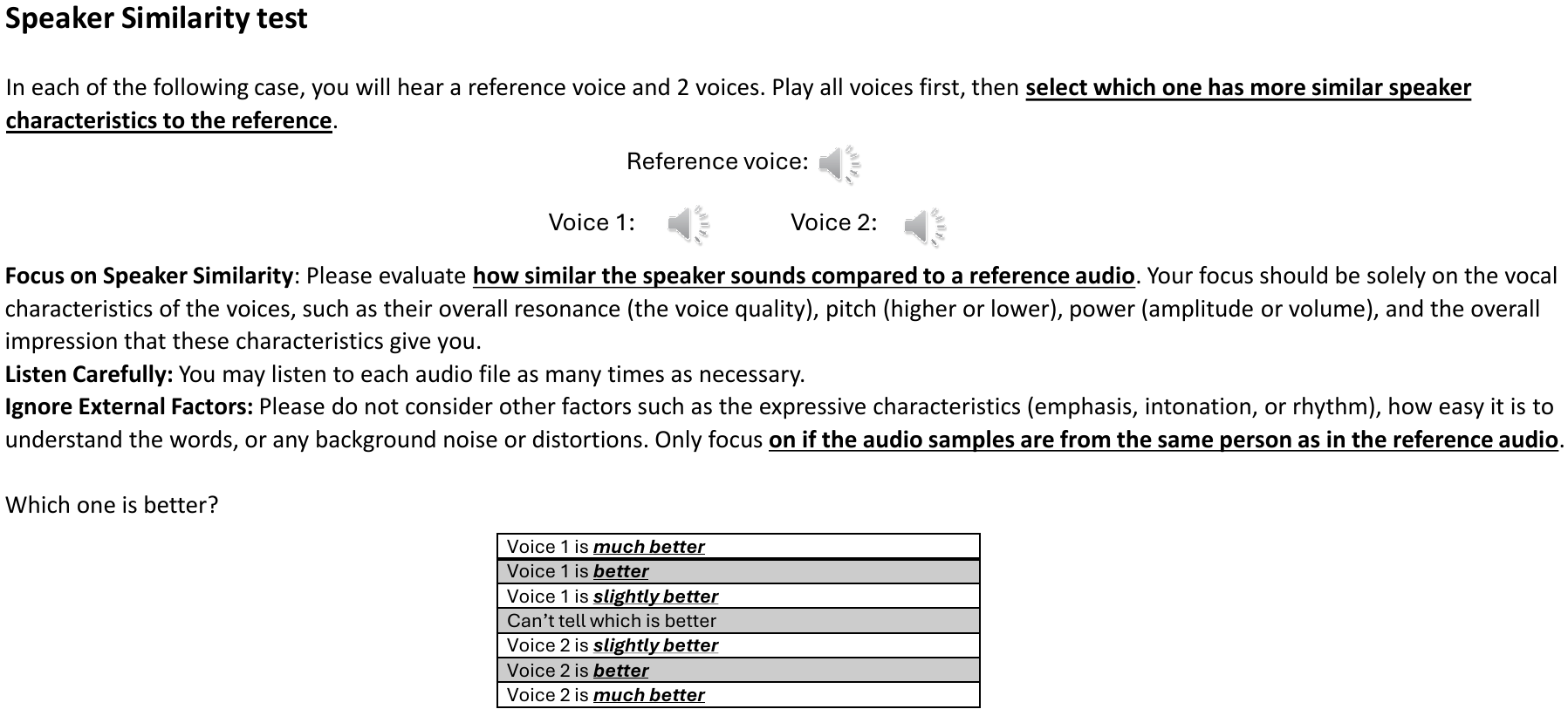}}
\caption{SMOS evaluation template}
\label{fig:mono-smos}
\end{figure}

Figure \ref{fig:mono-cmos} and Figure  \ref{fig:dia-cmos} shows the CMOS subjective evaluation template for monologue and dialogue generation respectively. Figure \ref{fig:mono-smos} show the SMOS subjective evaluation template. 

\newpage
\section*{NeurIPS Paper Checklist}

\begin{enumerate}

\item {\bf Claims}
    \item[] Question: Do the main claims made in the abstract and introduction accurately reflect the paper's contributions and scope?
    \item[] Answer: \answerYes{} 
    \item[] Justification: We introduce CoVoMix, a novel model for zero-shot, human-like, multi-speaker, multi-round dialogue speech generation. Our experimental results show that CoVoMix can generate dialogues that are not only human-like in their naturalness and coherence but also involve multiple talkers engaging in multiple rounds of conversation.


\item {\bf Limitations}
    \item[] Question: Does the paper discuss the limitations of the work performed by the authors?
    \item[] Answer: \answerYes{} 
    \item[] Justification: In the last section, we mentioned our limitation. We have observed instances of words being omitted or duplicated occasionally in synthesized speech and the use of low-quality dataset may degrade the quality of generated speech. 

\item {\bf Theory Assumptions and Proofs}
    \item[] Question: For each theoretical result, does the paper provide the full set of assumptions and a complete (and correct) proof?
    \item[] Answer: \answerNA{} 
    \item[] Justification: This paper mainly focused on the design of generation pipeline, model architecture, data preparation, and system implementation. We did not propose new theoretical algorithms or results. 

    \item {\bf Experimental Result Reproducibility}
    \item[] Question: Does the paper fully disclose all the information needed to reproduce the main experimental results of the paper to the extent that it affects the main claims and/or conclusions of the paper (regardless of whether the code and data are provided or not)?
    \item[] Answer: \answerYes{} 
    \item[] Justification: We provide detailed introduction of system design, data preparation and training configuration. We use public available dataset, provide all codes in the Supplementary and will make them publicly available.

\item {\bf Open access to data and code}
    \item[] Question: Does the paper provide open access to the data and code, with sufficient instructions to faithfully reproduce the main experimental results, as described in supplemental material?
    \item[] Answer: \answerYes{} 
    \item[] Justification: We use public available dataset. We provide detailed codes of training, inference and data preparation in the Supplementary. These codes will be publicly available.

\item {\bf Experimental Setting/Details}
    \item[] Question: Does the paper specify all the training and test details (e.g., data splits, hyperparameters, how they were chosen, type of optimizer, etc.) necessary to understand the results?
    \item[] Answer: \answerYes{} 
    \item[] Justification: We introduce in detail the training and test configuration, including model architecture, optimizer and hyperparameters, etc. We also provide ablation studies for these configurations.  Corresponding codes are in Supplementary and will be publicly available.

\item {\bf Experiment Statistical Significance}
    \item[] Question: Does the paper report error bars suitably and correctly defined or other appropriate information about the statistical significance of the experiments?
    \item[] Answer: \answerYes{} 
    \item[] Justification: We demonstrate the experiment statistical significance in Appendix. We calculate the mean and standard deviation of each system with different random seeds. 

\item {\bf Experiments Compute Resources}
    \item[] Question: For each experiment, does the paper provide sufficient information on the computer resources (type of compute workers, memory, time of execution) needed to reproduce the experiments?
    \item[] Answer: \answerYes{} 
    \item[] Justification: We provide detailed configuration of training and test, including the GPU, memory, training epochs, evaluation tools, etc.
    
\item {\bf Code Of Ethics}
    \item[] Question: Does the research conducted in the paper conform, in every respect, with the NeurIPS Code of Ethics \url{https://neurips.cc/public/EthicsGuidelines}?
    \item[] Answer: \answerYes{} 
    \item[] Justification: This research conducted in the paper conforms with the NeurIPS Code of Ethics. 

\item {\bf Broader Impacts}
    \item[] Question: Does the paper discuss both potential positive societal impacts and negative societal impacts of the work performed?
    \item[] Answer: \answerYes{} 
    \item[] Justification: We discussed broader impacts and potential risks in the last paragraph of Section 7.
    
\item {\bf Safeguards}
    \item[] Question: Does the paper describe safeguards that have been put in place for responsible release of data or models that have a high risk for misuse (e.g., pretrained language models, image generators, or scraped datasets)?
    \item[] Answer: \answerNA{}
    \item[] Justification: Currently we do not have plans for releasing the model or dataset. If we are going to release, we will make sure to release a safeguard with it.

\item {\bf Licenses for existing assets}
    \item[] Question: Are the creators or original owners of assets (e.g., code, data, models), used in the paper, properly credited and are the license and terms of use explicitly mentioned and properly respected?
    \item[] Answer: \answerYes{} 
    \item[] Justification: All the open-source code and evaluation tools that we use are credited in this paper, and the license and terms are properly respected. 

\item {\bf New Assets}
    \item[] Question: Are new assets introduced in the paper well documented and is the documentation provided alongside the assets?
    \item[] Answer: \answerYes{} 
    \item[] Justification: We introduce the implementation code as a new asset. This asset is well documented with training, license, limitations, etc. 

\item {\bf Crowdsourcing and Research with Human Subjects}
    \item[] Question: For crowdsourcing experiments and research with human subjects, does the paper include the full text of instructions given to participants and screenshots, if applicable, as well as details about compensation (if any)? 
    \item[] Answer: \answerNA{} 
    \item[] Justification: Our paper neither involves crowdsourcing nor research with human subjects.

\item {\bf Institutional Review Board (IRB) Approvals or Equivalent for Research with Human Subjects}
    \item[] Question: Does the paper describe potential risks incurred by study participants, whether such risks were disclosed to the subjects, and whether Institutional Review Board (IRB) approvals (or an equivalent approval/review based on the requirements of your country or institution) were obtained?
    \item[] Answer: \answerNA{}
    \item[] Justification: Our paper neither involves crowdsourcing nor research with human subjects.

\end{enumerate}

\end{document}